\documentclass[preprintnumbers,prd,amsmath,amssymb,nofootinbib,superscriptaddress]{revtex4}

\usepackage{graphicx}
\usepackage{bm}
\newcommand{\beq}{\begin{equation}}
\newcommand{\eeq}{\end{equation}}
\newcommand{\bea}{\begin{eqnarray}}
\newcommand{\eea}{\end{eqnarray}}

\newcommand{\Ref}[1]{(\ref{#1})}

\begin{document}

\title{Scalar multi-wormholes}
\author{A. I. Egorov$^1$, P. E. Kashargin$^1$, and Sergey V. Sushkov}
\email{sergey_sushkov@mail.ru}
\affiliation{Institute of Physics, Kazan Federal University, \\
Kremlevskaya 16a, Kazan 420008, Russia}

\begin{abstract}
In 1921 Bach and Weyl derived the method of superposition to construct new axially symmetric vacuum solutions of General Relativity. In this paper we extend the Bach-Weyl approach to non-vacuum configurations with massless scalar fields. Considering a phantom scalar field with the negative kinetic energy, we construct a multi-wormhole solution describing an axially symmetric superposition of $N$ wormholes. The solution found is static, everywhere regular and has
no event horizons. These features drastically tell the multi-wormhole configuration from other axially symmetric vacuum solutions which inevitably contain gravitationally inert singular structures, such as `struts' and `membranes', that keep the two bodies apart making a stable configuration. However, the multi-wormholes are static without any singular struts. Instead, the stationarity of the multi-wormhole configuration is provided by the phantom scalar field with the negative kinetic energy. Anther unusual property is that the multi-wormhole spacetime has a complicated topological structure. Namely, in the spacetime there exist $2^N$ asymptotically flat regions connected by throats.
\end{abstract}

 \maketitle

\section{Introduction}
The two-body (and, more generally, $N$-body) problem is one of the most
important and interesting to be solved in any theory of gravity. It is well
known that in Newtonian gravity this problem is well-defined and, in the
two-body case $N = 2$, has a simple exact treatment of the solution.
In general relativity the $N$-body problem has been investigated from the
early days of Einstein's gravitation theory
\cite{Dro,DeSitter,LorDro,LC,EinInfHof}. However, because of conceptual and
technical difficulties the motion of the $N$ bodies cannot be solved exactly in
the context of general relativity, even when $N = 2$. Hence analysis of a two
body system (e.g. binary pulsars) necessarily involves resorting to
approximation methods such as a post-Newtonian expansion (see, e.g.,
\cite{Damour, Blanchet} for a review and references to the early literature).

Besides developing approximate methods for solving the $N$-body problem,
essential efforts had been undertaken to search and study exact solutions of
the Einstein equations which could be interpreted as that describing
multi-particle configurations in general relativity. At the early days of general relativity
Weyl \cite{Weyl} and Levi-Civita \cite{LeviCivita} demonstrated that static axially symmetric gravitational fields can be described by the following metric (now known as the Weyl metric):
\beq\label{axmetric}
ds^2=-e^{2\lambda}dt^2+e^{-2\lambda}\left[e^{2\nu}
(d\rho^2+dz^2)+\rho^2d\varphi^2\right],
\eeq
where the functions $\lambda$ and $\nu$ depend on coordinates $\rho$ and $z$ only. In vacuum, when $R_{\mu\nu}=0$, these functions satisfy, respectively, the Laplace equation
\beq
\label{vfeq1}
\Delta\lambda=\lambda_{,\rho\rho}+\lambda_{,zz}+\frac{\lambda_{,\rho}}{\rho}=0,
\eeq
and two partial differential equations
\bea
&&\label{vfeq2}
\nu_{,\rho}=\rho(\lambda_{,\rho}^2-\lambda_{,z}^2),\\
&&\label{vfeq3}
\nu_{,z}=2\rho\lambda_{,\rho}\lambda_{,z},
\eea
where a comma denotes a partial derivative, e.g. $\lambda_{,\rho}={\partial\lambda}/{\partial\rho}$, $\lambda_{,\rho\rho}={\partial^2\lambda}/{\partial\rho^2}$, etc.
Since \Ref{vfeq1} is linear and homogeneous, the superposition of any integrals is again an integral of that equation. This fact and the particular form of Eqs. \Ref{vfeq2}, \Ref{vfeq3} allows us to find explicitly solutions that represent the superposition of two or more axially symmetric bodies of equal or different shapes.
In the pioneering work \cite{BachWeyl}, Bach and Weyl obtained an axially symmetric vacuum solution interpreted as
equilibrium configurations of a pair of Schwarzschild black holes. Since then, the fruitful Bach-Weyl method
has been used to construct a number of exact axially symmetric vacuum solutions \cite{Chazy,Curzon,Chou,Synge,Bondi,IsraelKhan,Letelier:87,Letelier:88}.
It is worth noticing that all solutions, representing the superposition of two or more axially symmetric bodies, contain gravitationally inert singular structures, `struts' and `membranes', that keep the two bodies apart making a stable configuration. This is an expected ``price to pay'' for the stationarity \cite{BachWeyl,EinsteinRosen,Cooperstock,Tomimatsu,Yamazaki,Schleifer}.

The advances of the Bach-Weyl approach in constructing exact axially symmetric solutions are based on the specific algebraic form of the {\em vacuum} field equations \Ref{vfeq1}-\Ref{vfeq3}. However, in this paper we demonstrate that this approach can be easily extended to gravitating systems with a massless scalar field $\phi$. Though such the systems are not vacuum ones, we show that new axially symmetric solutions with the scalar field can be found as a superposition of known spherically symmetric configurations. More specifically, we construct and analyze a multi-wormhole configuration with the {phantom} scalar field that possesses the negative kinetic energy.

The paper is organized as follows.
In the section \ref{sec_fe} we consider the theory of gravity with a massless scalar field and derive field equations for a static axially symmetric configuration. In the section \ref{sec_wh} we review solutions describing spherically symmetric scalar wormholes. In the section \ref{sec_multiwh} we construct multi-wormhole solutions. First, we consider specific properties of a single-wormhole axially symmetric configuration. Then, using the superposition method, we discuss in detail the procedure of constructing of two-wormhole solution, and, more generally, $N$-wormhole one. In the concluding section we summarize results and give some remarks concerning the regularity of the solutions and the violation of the null energy condition.

%--------------------------------------------------------------------
\section{Field equations\label{sec_fe}}
%--------------------------------------------------------------------
The theory of gravity with a massless scalar field $\phi$ is described by the action
\begin{equation}\label{action}
S=\int d^4x\sqrt{-g}\left[ R - \epsilon g^{\mu\nu}\phi_{,\mu}\phi_{,\nu}\right],
\end{equation}
where $g_{\mu\nu}$ is a metric, $g=\det(g_{\mu\nu})$, and $R$ is the scalar curvature.
The parameter $\epsilon$ equals $\pm1$. In the case $\epsilon=1$ we have a canonical scalar field with the positive kinetic term, and the case $\epsilon=-1$ describes a {\em phantom} scalar field with the negative kinetic energy. It is well-known that wormholes in the theory \Ref{action} exist only if the scalar field is phantom, and so hereinafter we will assume that $\epsilon=-1$.

Varying the action \Ref{action} with respect to $g_{\mu\nu}$ and $\phi$ leads to the field equations
\begin{subequations} \label{fieldeq}
\beq\label{eineq}
R_{\mu\nu}=-\phi_{,\mu}\phi_{,\nu},
\eeq
\beq \label{eqmo}
\square\phi=0.
\eeq
\end{subequations}

Let us search for static axially symmetric solutions of Eqs. \Ref{fieldeq}. In this case a gravitational field is described by the canonical Weyl metric \Ref{axmetric}, and the scalar field $\phi$ depends on coordinates $\rho$ and $z$ only. Now the field equations \Ref{fieldeq} can be represented in the following form
\begin{subequations}\label{feq}
\bea
&&\label{feq4}
\Delta\lambda=0,\\
&&\label{feq5}
\Delta\phi=0, \\
&&\label{feq2}
\nu_{,\rho}=\rho\left[\lambda_{,\rho}^2-\lambda_{,z}^2-\textstyle\frac12(\phi_{,
\rho }
^2-\phi_{,z}^2)\right],\\
&&\label{feq3}
\nu_{,z}=\rho\left[2\lambda_{,\rho}\lambda_{,z}-\phi_{,\rho}\phi_{,
z }\right].
\eea
\end{subequations}
Here it is necessary to emphasize that the functions $\lambda(\rho,z)$ and $\phi(\rho,z)$ satisfy the Laplace equations \Ref{feq4}, \Ref{feq5}, where
$\textstyle \Delta=\frac{1}{\rho}\frac{\partial}{\partial\rho}\left(\rho\frac{\partial}{\partial\rho}\right)+\frac{\partial^2}{
\partial z^2}$ is the Laplace operator in cylindrical coordinates. Since \Ref{feq4}, \Ref{feq5} are linear and homogeneous, superpositions of any integrals are again integrals of that equations.
Then, since $\nu_{,\rho z}=\nu_{,z\rho}$, Eqs. \Ref{feq2}, \Ref{feq3} yield
$$
2\lambda_{,z} \Delta\lambda+ \phi_{,z}\Delta\phi=0.
$$
The latter is fulfilled due to \Ref{feq4} and \Ref{feq5}, and so Eqs.
\Ref{feq4}, \Ref{feq5} are integrability conditions for Eqs.
\Ref{feq2}, \Ref{feq3}. Therefore, provided solutions $\lambda(\rho,z)$ and
$\phi(\rho,z)$ of Eqs. \Ref{feq4}, \Ref{feq5}  are given, Eqs.
\Ref{feq2}, \Ref{feq3} can be integrated in terms of the line integral
\beq\label{lineintegral}
\nu(\rho,z)=\int_{\cal C} \rho \big[\lambda_{,\rho}^2-\lambda_{,z}^2
-{\textstyle\frac12}(\phi_{,\rho}^2-\phi_{,z}^2)\big]d\rho
+ \rho \big[2\lambda_{,\rho}\lambda_{,z}-\phi_{,\rho}\phi_{,z}\big]
dz,
\eeq
where $\cal C$ is an arbitrary path connecting some fixed point
$(\rho_0,z_0)$ with the point $(\rho,z)$. In practice, a choice of
$(\rho_0,z_0)$ is determined by the boundary condition $\nu(\rho_0,z_0)=\nu_0$.

%%%%%%%%%%%%%%%%%%%%%%%%%%%%%%%%%%%%%%%%%%%%%%%%%%%%
\section{Spherically symmetric wormhole solution \label{sec_wh}}
%%%%%%%%%%%%%%%%%%%%%%%%%%%%%%%%%%%%%%%%%%%%%%%%%%%%
Wormholes are usually defined as topological handles in spacetime linking widely separated regions of a single universe, or ``bridges'' joining two different spacetimes \cite{MorTho,VisserBook}. As is well-known \cite{HocVis1,HocVis2}, in the framework of general relativity they can exist only if their throats contain an exotic matter which possesses a negative pressure and violates the null energy condition. The simplest model which provides the necessary conditions for existence of wormholes is the theory of gravity \Ref{action} with the phantom scalar field. Static spherically symmetric wormholes in the model \Ref{action} were obtained
by Ellis \cite{Ellis} and Bronnikov \cite{Bro_73}. Adapting their result (see Ref. \cite{SusKim}), one can write down a wormhole solution as follows
\beq\label{ssswhmetric0}
ds^2=-e^{2\lambda(r)}dt^2+e^{-2\lambda(r)}\left[dr^2+(r^2+a^2)\left (
d\theta^2+\sin^2\theta d\varphi^2\right) \right],
\eeq
\beq\label{ssswhsf0}
\phi(r)=\frac{\sqrt{2(m^2+a^2)}}{m}\,\lambda(r),
\eeq
where
\beq
\lambda(r)=\frac{m}{a}\arctan({r}/{a}),
\eeq
$r\in(-\infty,\infty)$, and $m$ and $a$ are two free parameters.
In the case $m=0$ one has $\lambda(r)=0$, and the solution
\Ref{ssswhmetric0}, \Ref{ssswhsf0} takes the especially simple form:
\beq
\label{ssswhmetric}
ds^2=-dt^2+dr^2+(r^2+a^2)\left(d\theta^2+\sin^2\theta d\varphi^2\right),
\eeq
\beq\label{ssswhsf}
\phi(r)=\sqrt{2}\,\arctan({r}/{a}).
\eeq
The latter describes a massless wormhole connecting two asymptotically flat
regions at $r\to\pm\infty$. The throat of the wormhole is located at $r=0$ and has
the radius $a$. The scalar field $\phi$ smoothly varies between two asymptotical
values $\pm\pi/\sqrt{2}$.

Hereafter, for simplicity, we will assume that $m=0$ and hence $\lambda=0$.

%%%%%%%%%%%%%%%%%%%%%%%%%%%%%%%%%%%%%%%%%%%%%%%%%%%%
\section{Multi-wormhole solution \label{sec_multiwh}}
%%%%%%%%%%%%%%%%%%%%%%%%%%%%%%%%%%%%%%%%%%%%%%%%%%%%
\subsection{Single wormhole} \label{Sec-swh}
%%%%%%%%%%%%%%%%%%%%%%%%%%%%%%%%%%%%%%%%%%%%%%%%%%%%

Let us consider an axially symmetric form of the spherically symmetric wormhole solution given in the previous section. For this aim we rewrite the solution \Ref{ssswhmetric}-\Ref{ssswhsf} in cylindrical coordinates $(\rho,z)$
carrying out the coordinate transformation
\begin{subequations} \label{ctrans}
\bea
\rho&=&\sqrt{r^2+a^2}\sin\theta,\\
z&=&r\cos\theta.
\eea
\end{subequations}
The corresponding Jacobian reads
\beq
J=\frac{D(\rho,z)}{D(r,\theta)}=-\frac{r^2+a^2\cos^2\theta}{\sqrt{r^2+a^2}}.
\eeq
Since the Jacobian $J$ is equal to zero at $r=0,\theta=\frac{\pi}{2}$, the transformation \Ref{ctrans} is degenerated at this point.

\begin{figure}
\includegraphics[width=8cm]{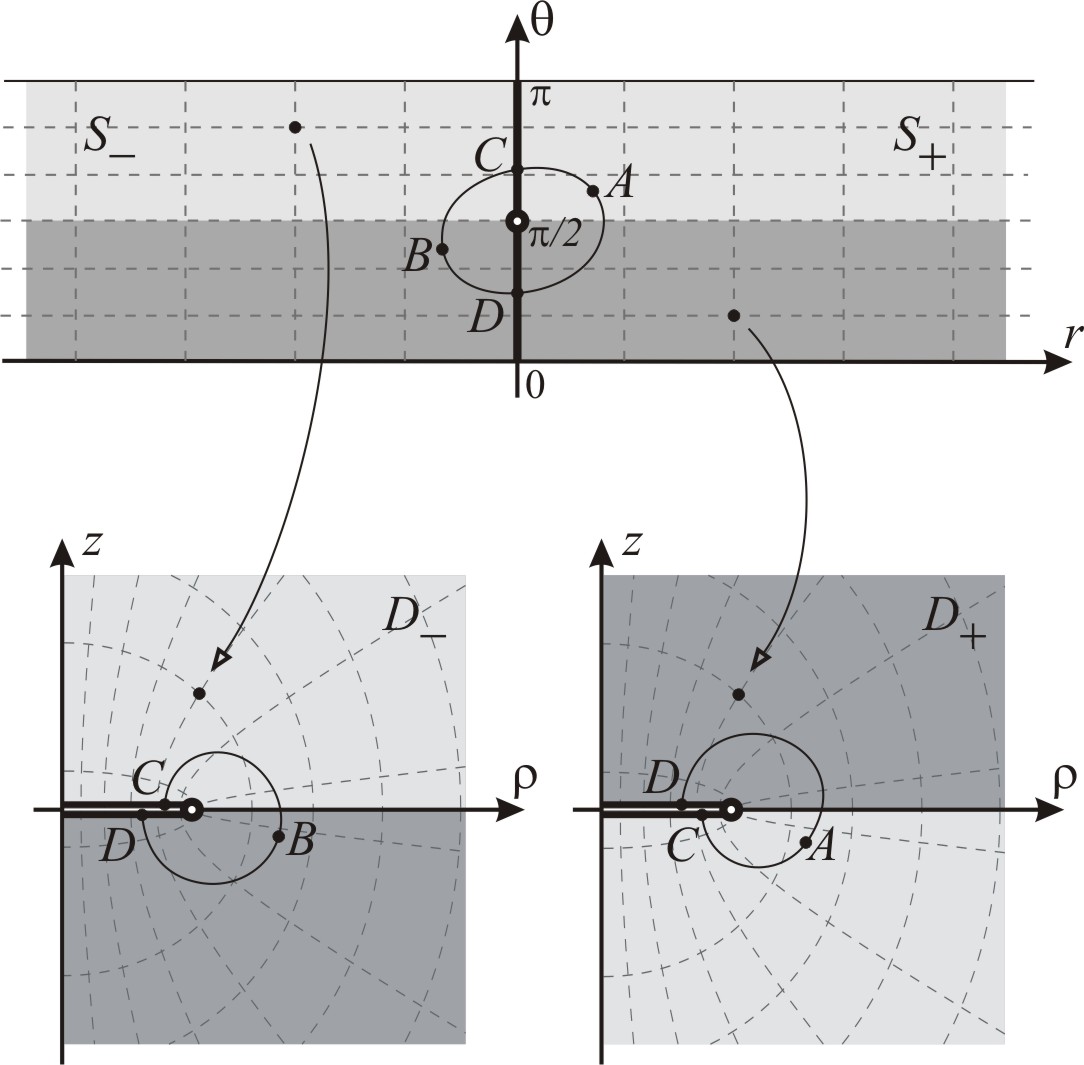}
\caption{\label{mapping} The mapping ${\cal S}\to \overline{\cal D}{}_2$ is shown. The regions ${\cal S}_+$ and ${\cal S}_-$ are mapped to ${\cal D}_+$ and ${\cal D}_-$  correspondingly. The segment ${\cal T}=\{r=0, \theta\in[0,\pi]\}$  in ${\cal S}$ corresponds to the wormhole's throat and it is mapped to two-sided segment ${\cal T}_{a\pm}=\{\rho\in[0,a], z=0\}$ in ${\cal D}_\pm$. Note that the upper part $\{r=0, \theta\in[\pi/2, \pi]\}$ of the throat ${\cal T}$ correspons to the upper part of the two-sided segment ${\cal T}_{a-}$ in ${\cal D}_-$ and the  lower part of the ${\cal T}_{a+}$ in ${\cal D}_+$, while the  lower part $\{r=0, \theta\in[0,\pi/2]\}$ of the throat ${\cal T}$ corresponds to the lower part of the two-sided segment ${\cal T}_{a -}$ of ${\cal D}_-$ and the upper part of the segment ${\cal T}_{a+}$ of ${\cal D}_+$.}
\end{figure}

Remind ourselves that in the wormhole spacetime the proper radial coordinate $r$ runs from $-\infty$ to $+\infty$, while the azimuthal coordinate $\theta$ runs from $0$ to $\pi$. Therefore, the domain of spherical coordinates $(r,\theta)$ can be represented as the strip ${\cal S} = (-\infty,\infty)\times[0,\pi]$ on the plane $\mathbb{R}\times\mathbb{R}$ (see Fig. \ref{mapping}). Note that the wormhole throat is represented as the segment
${\cal T}=\{r=0,\theta\in[0,\pi]\}$
on ${\cal S}$. The regions with $r\to\pm\infty$ on the strip $\cal S$ correspond to two asymptotically flat regions of the wormhole spacetime. The corresponding domain of cylindrical coordinates $(\rho,z)$ is found from \Ref{ctrans} as the half-plane ${\cal D}=[0,\infty)\times(-\infty,\infty)$. Correspondingly, the
wormhole throat is represented as the segment ${\cal T}_a=\{\rho\in[0,a], z=0\}$ on ${\cal D}$, where $a$ is the radius of the throat. Note also that the asymptotics $r\to\pm\infty$ in cylindrical coordinates take the form $(\rho^2+z^2)^{1/2}\to\infty$.

\begin{figure}
\includegraphics[width=8cm]{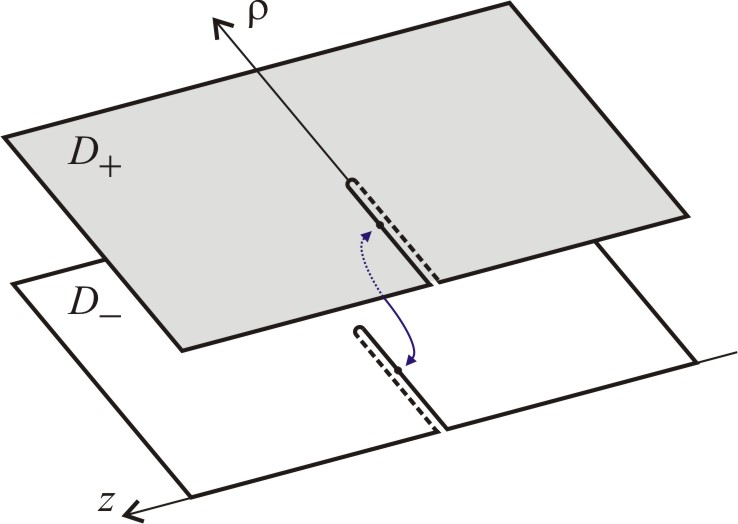}
\caption{\label{fig-D2} The manifold $\overline{\cal D}_2$ consists of two regions ${\cal D}_\pm$, which are glued together by the two-sided segments ${\cal T}_{a\pm}$.}
\end{figure}

The transformation \Ref{ctrans} determines the mapping ${\cal S}\to{\cal D}$. It is necessary to emphasize that it maps two different points of $\cal S$ into a single point of $\cal D$. Namely, $(r,\theta)\to(\rho,z)$ and $(-r,\pi-\theta)\to(\rho,z)$. Therefore, the mapping ${\cal S}\to{\cal D}$ is not bijective. To construct a
bijective mapping, we first consider two half-strips, ${\cal S}_-=(-\infty,0)\times[0,\pi]$ and ${\cal S}_+=(0,\infty)\times[0,\pi]$, which represent, respectively, `lower' ($r<0$) and `upper' ($r>0$) halves of the wormhole spacetime. The edge of both half-stripes is the segment $\cal T$, which, in turn, corresponds to the wormhole's throat. Additionally, we introduce two identical copies of the half-plane ${\cal D}$ denoted as ${\cal D}_\pm$; the segments ${\cal T}_{a\pm}$ on ${\cal D}_\pm$ correspond to the wormhole's throat. Now, each of the mapping ${\cal S}_\pm\to {\cal D}_\pm$ is bijective.
The whole strip $\cal S$ is restored by means of the gluing of two half-strips ${\cal S}_\pm$ along their common edge $\cal T$. This procedure corresponds to the topological gluing of two half-planes ${\cal D}_\pm$ by means of identifying corresponding points of the segments ${\cal T}_{a\pm}$.
As the result of this topological gluing, one obtains a new manifold denoted as
$\overline{\cal D}{}_2$, consisting of two sheets ${\cal D}_\pm$ with the common segment ${\cal T}_a$.
It is worth noticing that the manifold $\overline{\cal D}{}_2$ is homeomorphic to the self-crossing two-sheeted Riemann surface. The sheets ${\cal D}_\pm$ of the manifold $\overline{\cal D}{}_2$ contain asymptotically flat regions $(\rho^2+z^2)^{1/2}\to\infty$ connecting by the throat $\cal T$.
Now, the mapping ${\cal S}\to \overline{\cal D}{}_2$ is a bijection.
Schematically, the mapping ${\cal S}_\pm\to {\cal D}_\pm$ is shown in Fig.
\ref{mapping}. Respectively, in Fig. \ref{fig-D2} we illustrate the procedure of constructing
the manifold $\overline{\cal D}{}_2$ by means of topological gluing.

The transformation being inverse to \Ref{ctrans} reads
\begin{subequations} \label{inversetran}
\bea\label{rpm}
r&=&\pm R(\rho,z),
\\
\theta&=&\arccos \big({z}/{R(\rho,z)}\big),
\eea
\end{subequations}
where
\beq
R(\rho,z)=\frac{1}{\sqrt{2}}\left[(\rho^2+z^2-a^2)+
\sqrt{(\rho^2+z^2-a^2)^2+4a^2z^2}\right]^{1/2}.
\eeq
Performing the transformation \Ref{inversetran}, we obtain the single-wormhole metric \Ref{ssswhmetric} in Weyl coordinates:
\beq
ds_\pm^2=-dt^2+e^{2\nu_\pm}[d\rho^2+dz^2]
+\rho^2 d\varphi^2,
\eeq
where
\beq\label{nu1}
\nu_{\pm}(\rho,z)=\frac{1}{2}\ln\left[\frac{R^2(\rho,z)\big(R^2(\rho,
z)+a^2\big)}{R^4(\rho,z)+a^2z^2}\right],
\eeq
and the scalar field \Ref{ssswhsf} now reads
\beq\label{phi1}
\phi_{\pm}(\rho,z)=\pm\sqrt{2}\arctan \big({R(\rho,z)}/{a}\big).
\eeq
The signs $\pm$ in Eqs. \Ref{inversetran}-\Ref{phi1} are
pointing out that the metric $ds^2_\pm$, as well as the functions
$\phi_{\pm}(\rho,z)$ and $\nu_{\pm}(\rho,z)$ are defined on the respective
charts ${\cal D}_\pm$, i.e.
one should choose the plus sign if $(\rho,z)\in {\cal D}_+$ ($r>0$), and
minus if $(\rho,z)\in {\cal D}_-$ ($r<0$). The complete solutions defined on
the whole manifold $\overline{\cal D}{}_2$ are constituted from separate
$\phi_{\pm}(\rho,z)$ and $\nu_{\pm}(\rho,z)$. In particular, the scalar field
$\phi(\rho,z)$ on $\overline{\cal D}{}_2$ is defined as follows:
\beq\label{sol1}
\phi(\rho,z)=\left\{
\begin{array}{ll}
 \phi_-(\rho,z) & \mathrm{if}\ (\rho,z)\in {\cal D}_-\\
 \phi_+(\rho,z) & \mathrm{if}\ (\rho,z)\in {\cal D}_+
\end{array}
\right.
\eeq
\begin{figure}
\includegraphics[width=5cm]{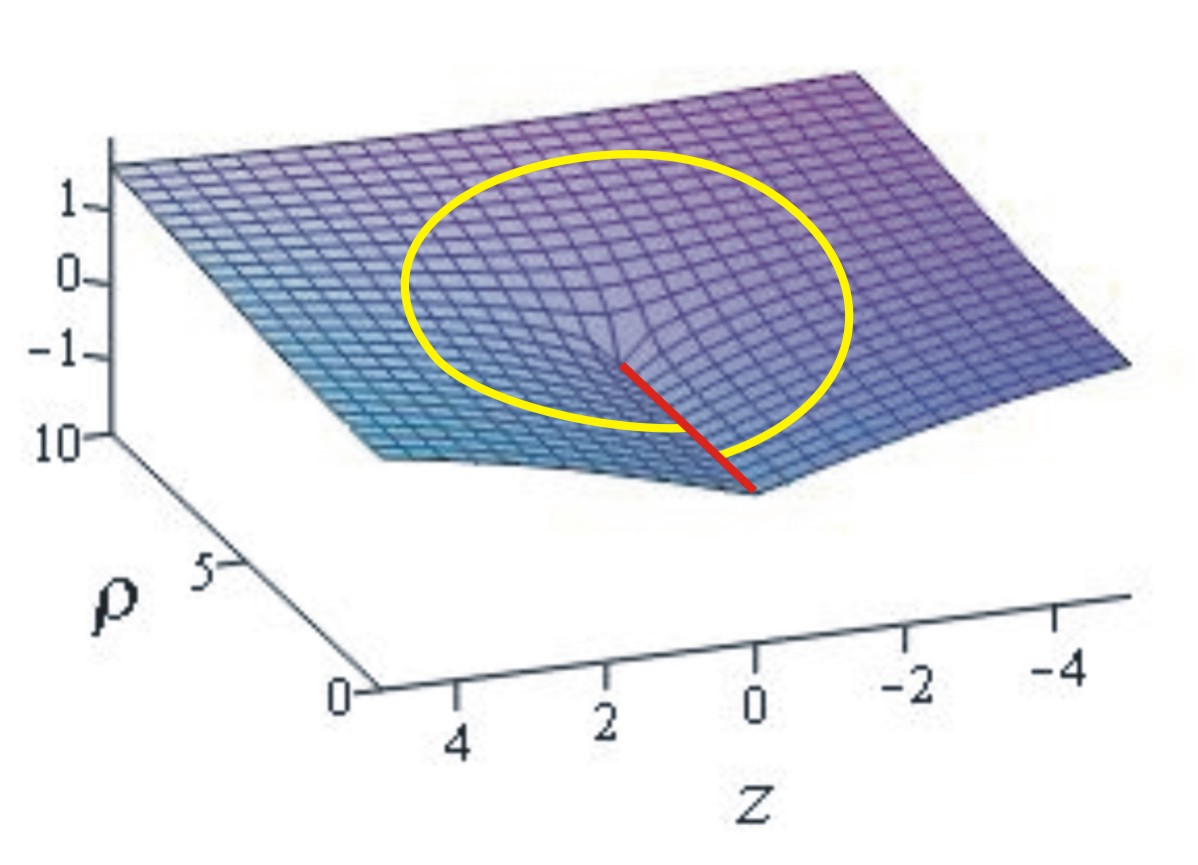}\
\includegraphics[width=5cm]{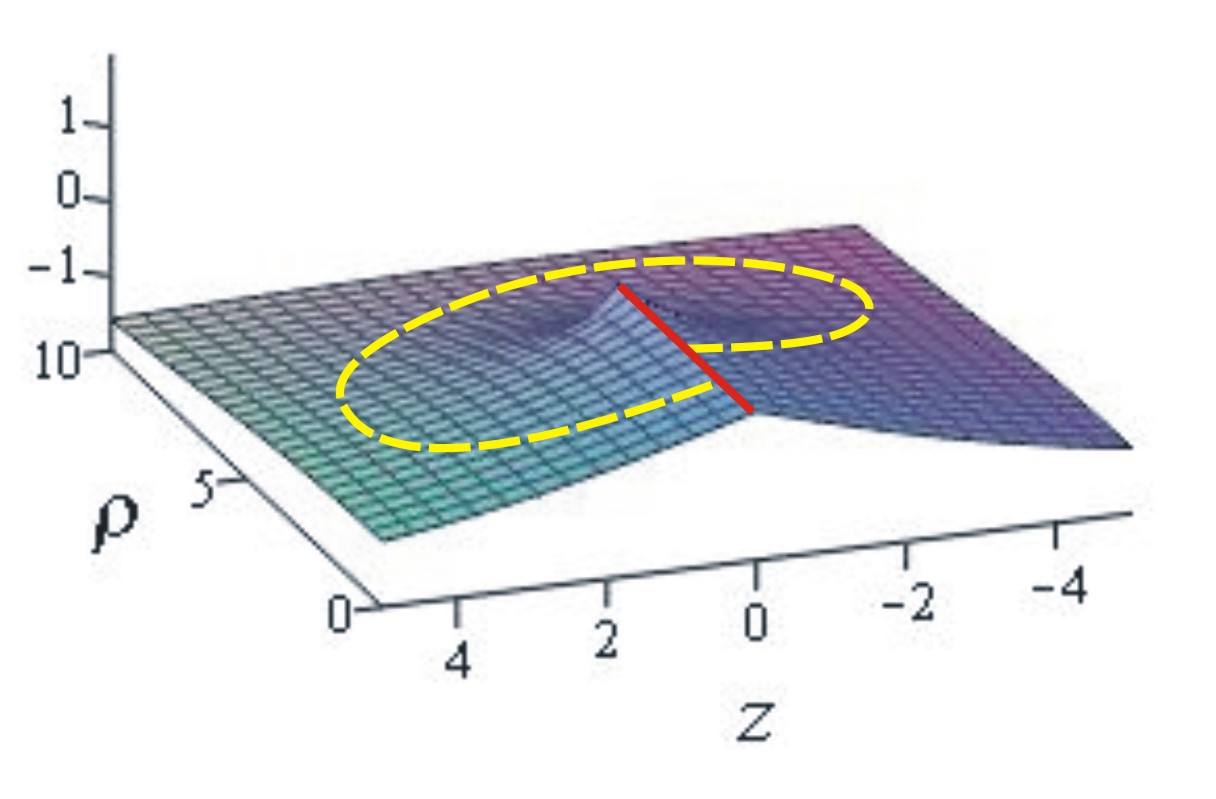}\
\caption{\label{fig-phipm} The plots of $\phi_{\pm}(\rho,z)$ defined on the respective
charts ${\cal D}_\pm$. The segments ${\cal T}_{a\pm}=\{\rho\in[0,a], z=0\}$ correspond to the wormhole's throat with the radius $a=5$.}
\end{figure}
Separately graphs of $\phi_{\pm}(\rho,z)$ are shown in Fig. \ref{fig-phipm}.
It is seen that $\phi_\pm(\rho,z)$ are not smooth at the segments ${\cal T}_{a\pm}$, i.e. at the throat. However, the complete solution $\phi(\rho,z)$, shown in Fig. \ref{fig-D2}, is single-valued, smooth and
differentiable everywhere on $\overline{\cal D}{}_2$.\footnote{Actually, it
is obvious that the scalar field, given in the
spherical coordinates as $\phi(r)=\sqrt{2}\arctan(r/a)$ (see Eq.\Ref{ssswhsf}),
is smooth and differentiable for all values of $r$ including the throat $r=0$.
Therefore, the scalar field \Ref{sol1} obtained from \Ref{ssswhsf} as the result
of coordinate transformations is also smooth and differentiable everywhere on
$\overline{\cal D}{}_2$.}

%%%%%%%%%%%%%%%%%%%%%%%%%%%%%%%%%%%%%%%%%%%%%%%%
\begin{figure}
\includegraphics[width=8cm]{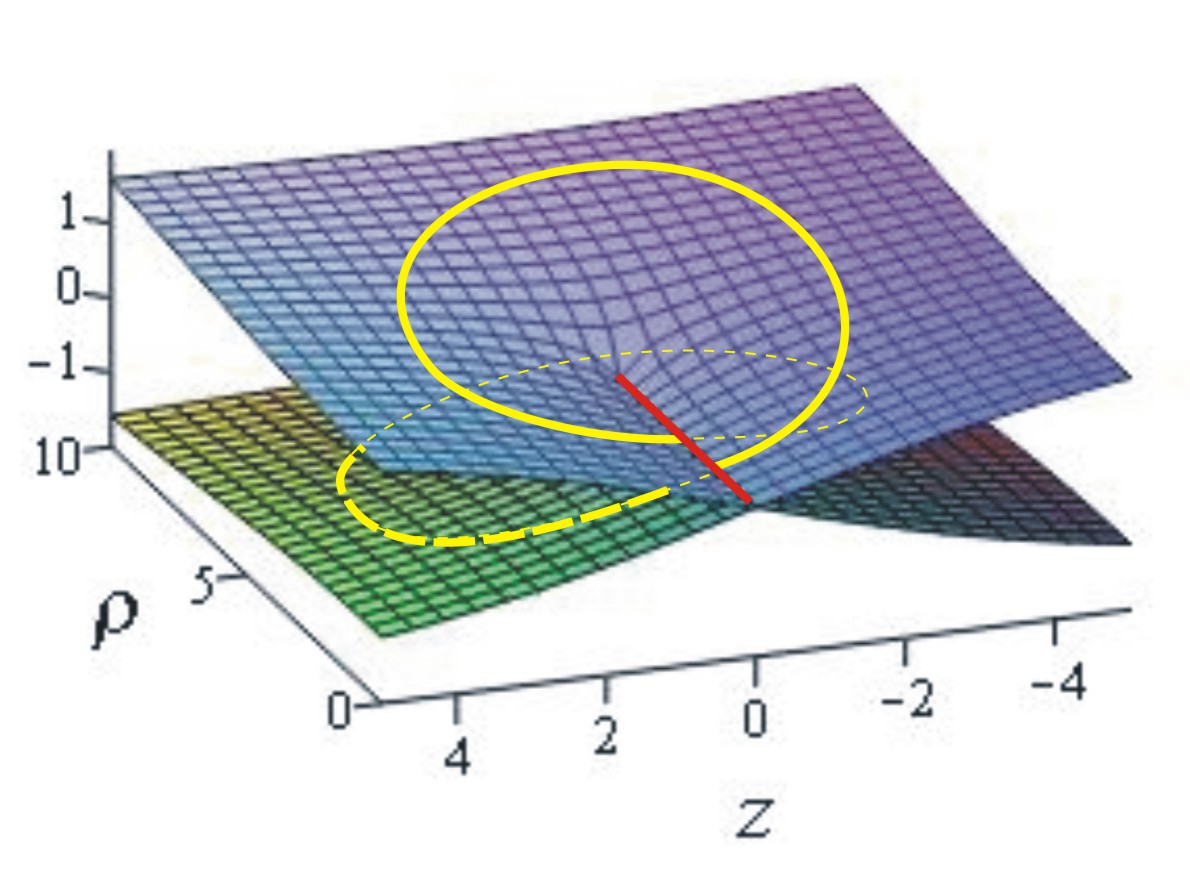}
\caption{\label{fig-phi1} The scalar field $\phi(\rho,z)$ on the manifold $\overline{\cal D}{}_2$ corresponding to the single-wormhole spacetime. The segment ${\cal T}=\{\rho\in[0,a], z=0\}$ corresponds to the wormhole's throat with the radius $a=5$.}
\end{figure}
%%%%%%%%%%%%%%%%%%%%%%%%%%%%%%%%%%%%%%%%%%%%%%%%

\begin{figure}
\includegraphics[width=6cm]{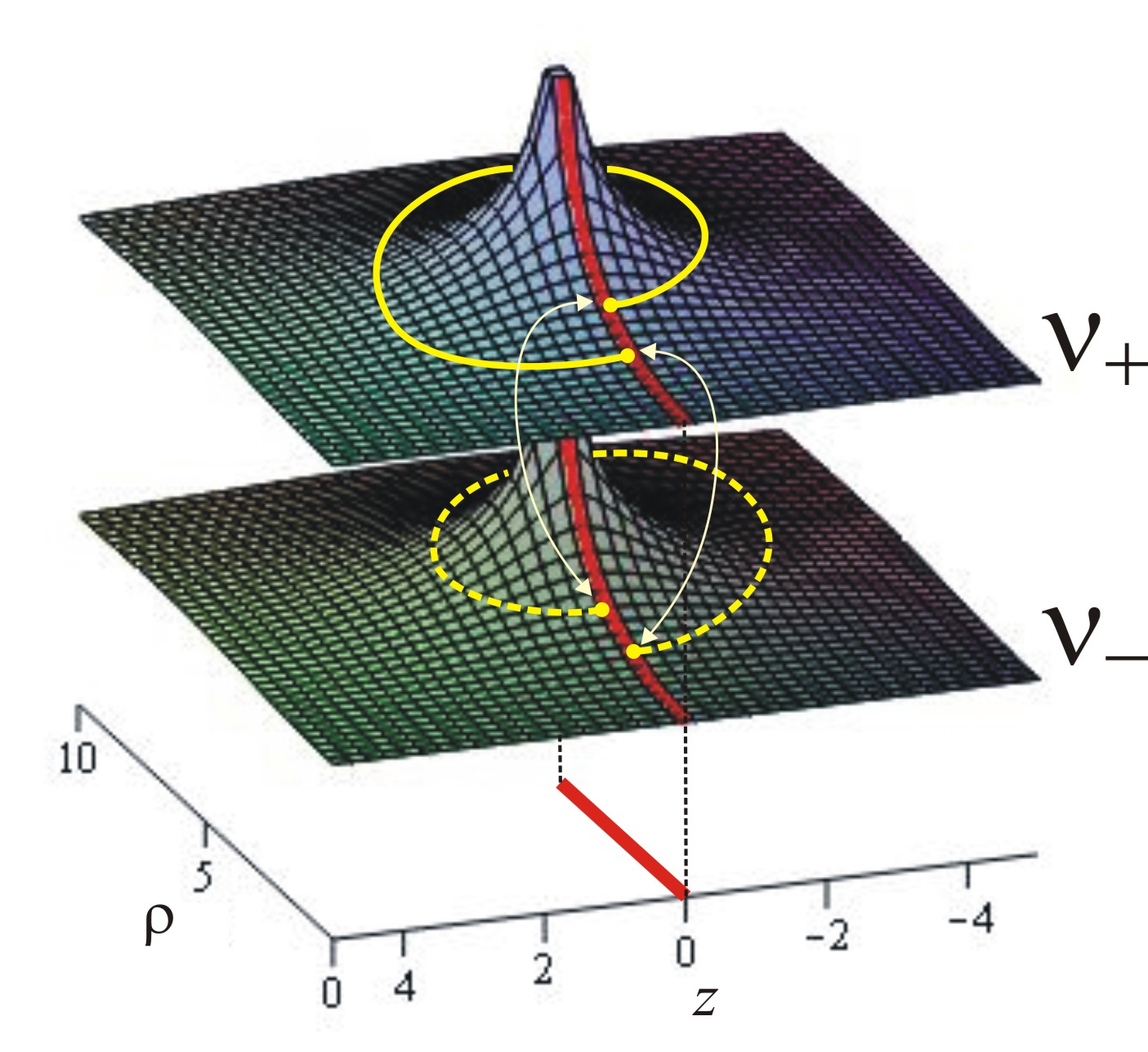}
\caption{\label{fig-nu1} A plot of $\nu(\rho,z)$. It consists of two identical sheets, $\nu_-(\rho,z)$ and $\nu_+(\rho,z)$,
topologically glued along the cuts on $\nu_\pm(\rho,z)$ with $\rho\in[0,a]$ and
$z=0$.}
\end{figure}

The metric function $\nu_\pm(\rho,z)$ given by Eq. \Ref{nu1} has
been obtained straightforwardly through the coordinate transformation
\Ref{ctrans} applied to the solution \Ref{ssswhmetric}. On the other hand,
$\nu(\rho,z)$ could be found directly as the line integral \Ref{lineintegral}. Since $\lambda(r)=0$, Eq. \Ref{lineintegral} reads
\beq
\nu(\rho,z)=\int_{\cal C}
\left[-{\textstyle\frac12}\rho(\phi_{,\rho}^2-\phi_{,z}^2)d\rho -\rho \phi_{,\rho}\phi_{,z}dz\right],\label{nu_1wh}
\eeq
where $\cal C$ is an arbitrary path connecting some initial point
$(\rho_0,z_0)$ and any point $(\rho,z)$ in $\overline{\cal D}{}_2$. Note
that if $(\rho_0,z_0)$ and $(\rho,z)$ are lying in the different charts ${\cal
D}_-$ and ${\cal D}_+$, then the path $\cal C$ should go through the segment
$\{\rho\in[0,a],z=0\}$, which corresponds to the wormhole's throat.
The coordinates of the initial point $(\rho_0,z_0)$ are determined by the boundary conditions for function $\nu$. Imposing the following boundary conditions
$$
\lim\limits_{\rho\to\infty,z\to -\infty}\nu(\rho,z)=0,\quad (\rho,z)\in {\cal D_+},
$$
substituting $\phi$ from Eq. \Ref{sol1}, and integrating, we obtain the expression \Ref{nu1}.
%\footnote{Details of calculations are given in the appendix.}
Finally, the complete solution $\nu(\rho,z)$ defined on $\overline{\cal D}{}_2$ is constituted from the separate $\nu_\pm(\rho,z)$ as follows
\beq\label{sol-nu1}
\nu(\rho,z)=\left\{
\begin{array}{ll}
 \nu_-(\rho,z) & \mathrm{if}\ (\rho,z)\in {\cal D}_-\\
 \nu_+(\rho,z) & \mathrm{if}\ (\rho,z)\in {\cal D}_+
\end{array}
\right.
\eeq
Since $\nu_-(\rho,z)$ and $\nu_+(\rho,z)$ are given by the same expression
\Ref{nu1}, a graph of $\nu(\rho,z)$ is represented by two identical sheets
topologically glued along the cuts on $\nu_\pm(\rho,z)$ with $\rho\in[0,a]$ and
$z=0$ (see Fig. \ref{fig-nu1}).

%%%%%%%%%%%%%%%%%%%%%%%%%%%%%%%%%%%%%%%%%%%%%%%%%%%%
\subsection{Two wormholes}
%%%%%%%%%%%%%%%%%%%%%%%%%%%%%%%%%%%%%%%%%%%%%%%%%%%%
Now, let us construct a two-wormhole solution. For this aim we remind ourselves that any superposition $c_1\phi_1+c_2\phi_2$ of two particular solutions $\phi_{1}$ and $\phi_{2}$ is a solution of the Laplace equation.
Moreover, because of the axial symmetry, if $\phi(\rho,z)$ is a solution,
then $\phi(\rho,z+\mathrm{const})$ is also a solution of $\Delta\phi=0$, where $\Delta=\frac{1}{\rho}\frac{\partial}{\partial\rho}\left({\rho\frac{\partial}{\partial\rho}}\right)+\frac{\partial^2}{\partial z^2}$ is the Laplace operator in cylindrical coordinates.
Using these properties, we may construct new axially symmetric solutions as a superposition of single-wormhole solutions \Ref{phi1}. In particular, the superposition of two solutions reads
\beq\label{phi2}
\phi_{\epsilon_1\epsilon_2}(\rho,z)=\frac12\left[\phi_{\epsilon_1}(\rho,
z-z_1;a_1)+\phi_{\epsilon_2} (\rho,z-z_2;a_2) \right],
\eeq
with
\beq
\phi_{\epsilon_i}(\rho,z-z_i;a_i)=\epsilon_i\sqrt{2}\arctan
\left(\frac{R(\rho,z-z_i)}{a_i}\right),
\eeq
where $i=1,2$, and indices $\epsilon_i$ take two values, $+$ or $-$, i.e. $\epsilon_i=\pm$.
Each particular solution $\phi_{\epsilon_i}(\rho,z-z_i;a_i)$ describes a
single wormhole with the throat of the radius $a_i$ located at $z_i$. In case $z_1=z_2=0$ and $a_1=a_2$, Eq. \Ref{phi2} reproduces the single-wormhole solution \Ref{phi1}. As the
result, the superposition \Ref{phi2} gives us four new solutions
$\phi_{++}$, $\phi_{+-}$, $\phi_{-+}$,
and $\phi_{--}$, which could be interpreted as the scalar field in the wormhole
spacetime with two throats of radiuses $a_{1}$ and $a_2$ located at $z_{1}$ and
$z_2$, respectively.

\begin{figure}
\includegraphics[width=8cm]{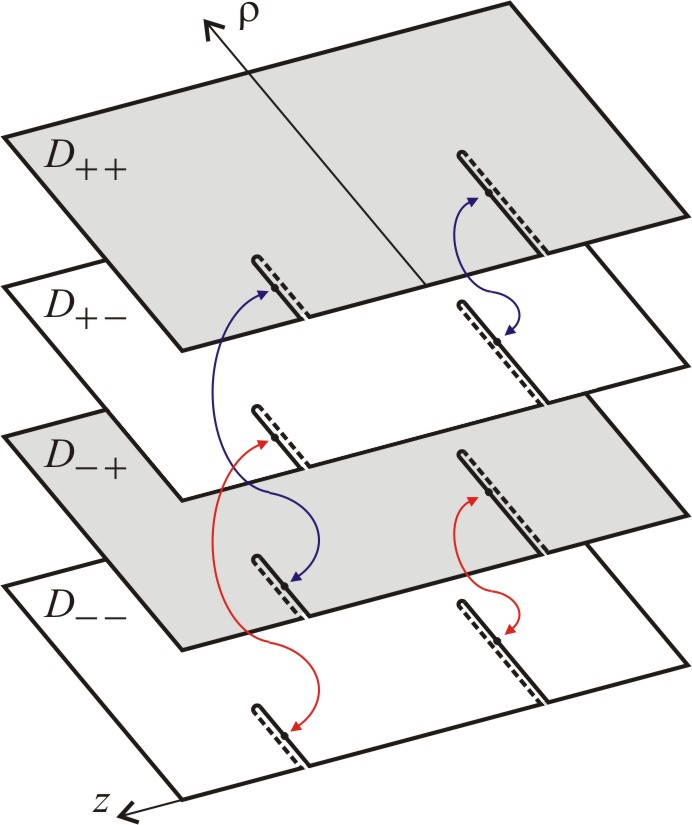}
\caption{\label{fig-D4}
The manifold $\overline{\cal D}{}_{4}$ consists of four identical copies of the
half-planes ${\cal D}_{\epsilon_1\epsilon_2}$ ($\epsilon_i=\pm$), which are glued together by the two-sided segments ${\cal T}_{a_i}=\{\rho\in[0,a_i],z=z_i\}$ ($i=1,2$) corresponding to two throats with radii $a_{i}$ located at $z_{i}$.}
\end{figure}

\begin{figure}
\includegraphics[width=4.0cm]{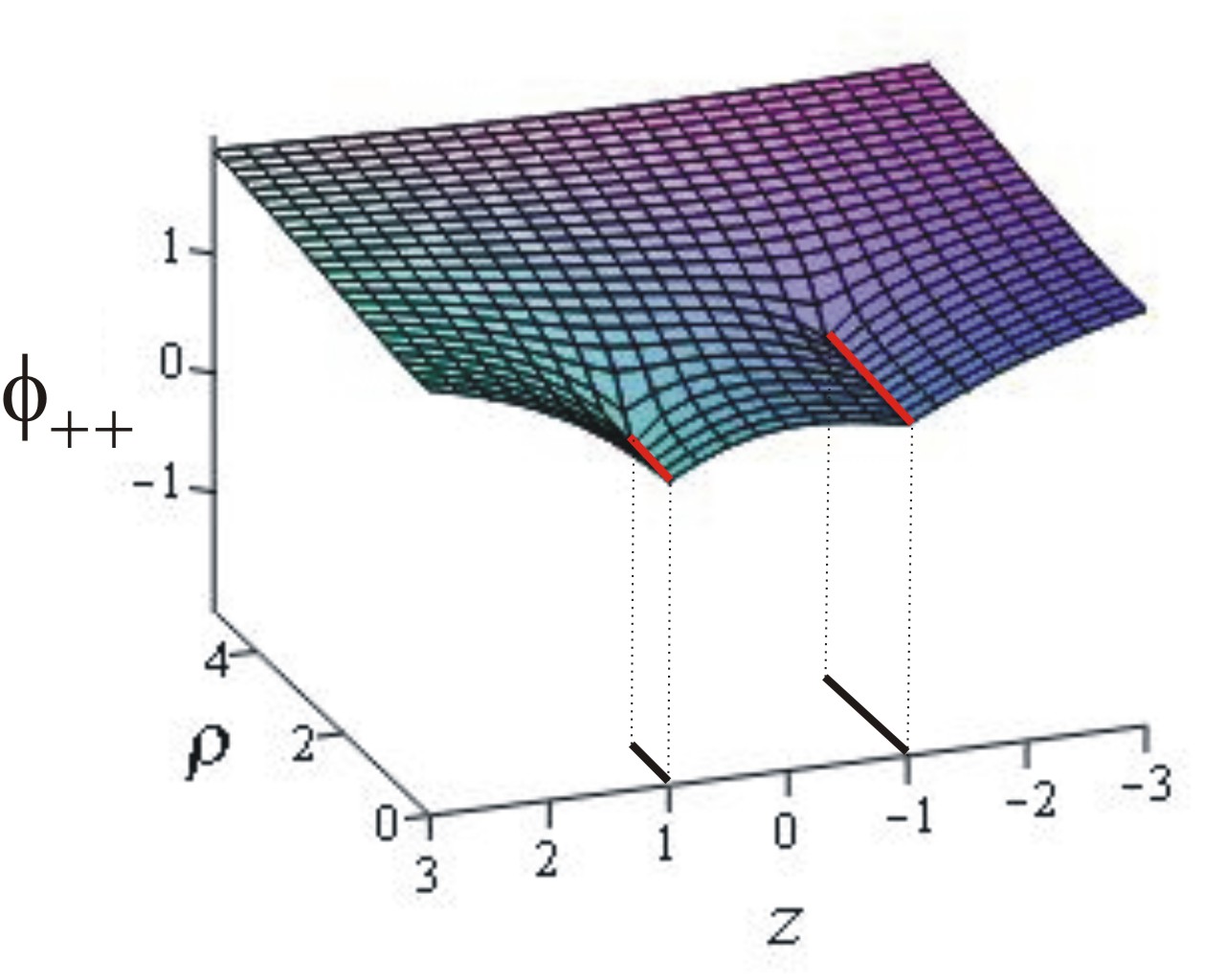}
\includegraphics[width=4.0cm]{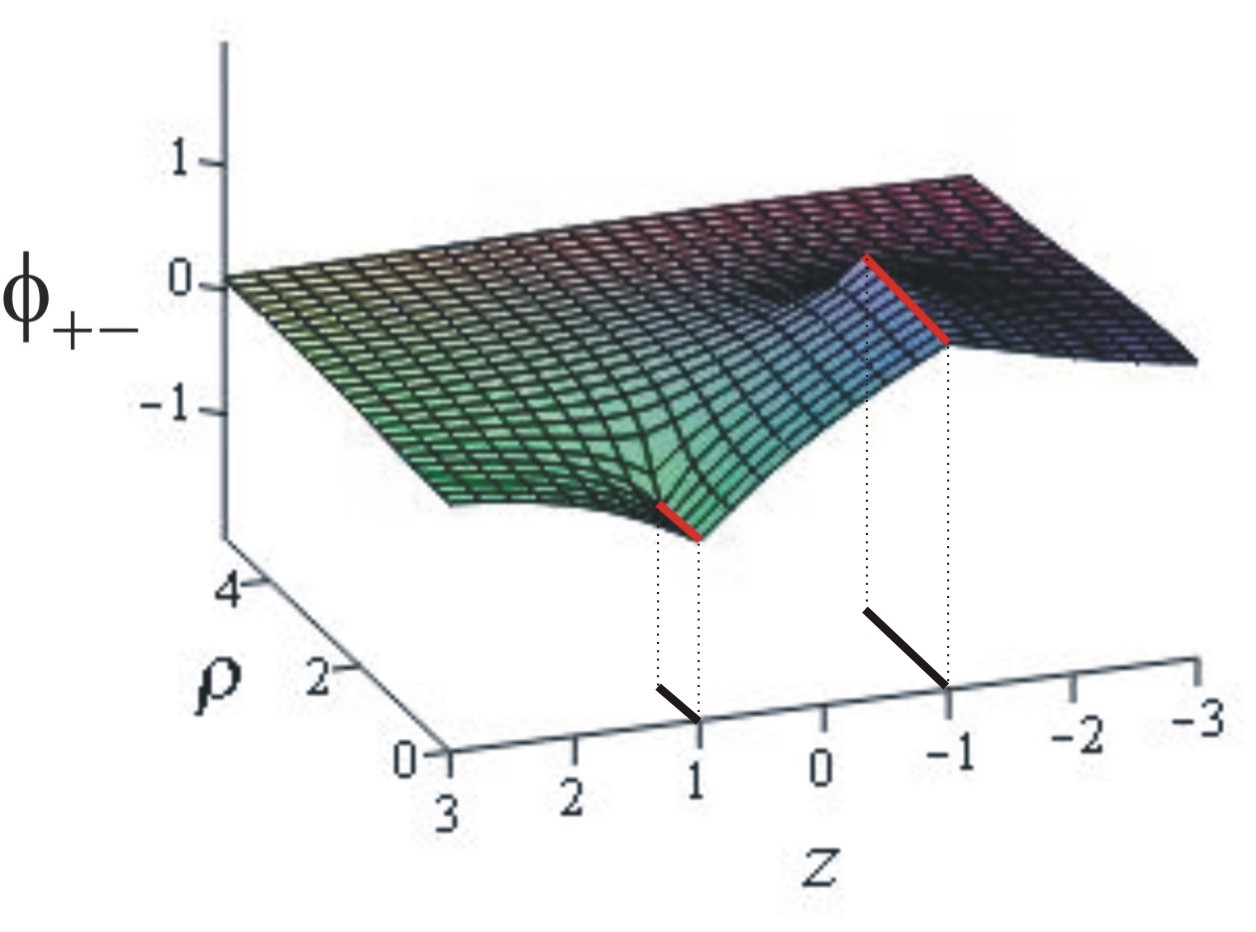}
\includegraphics[width=4.0cm]{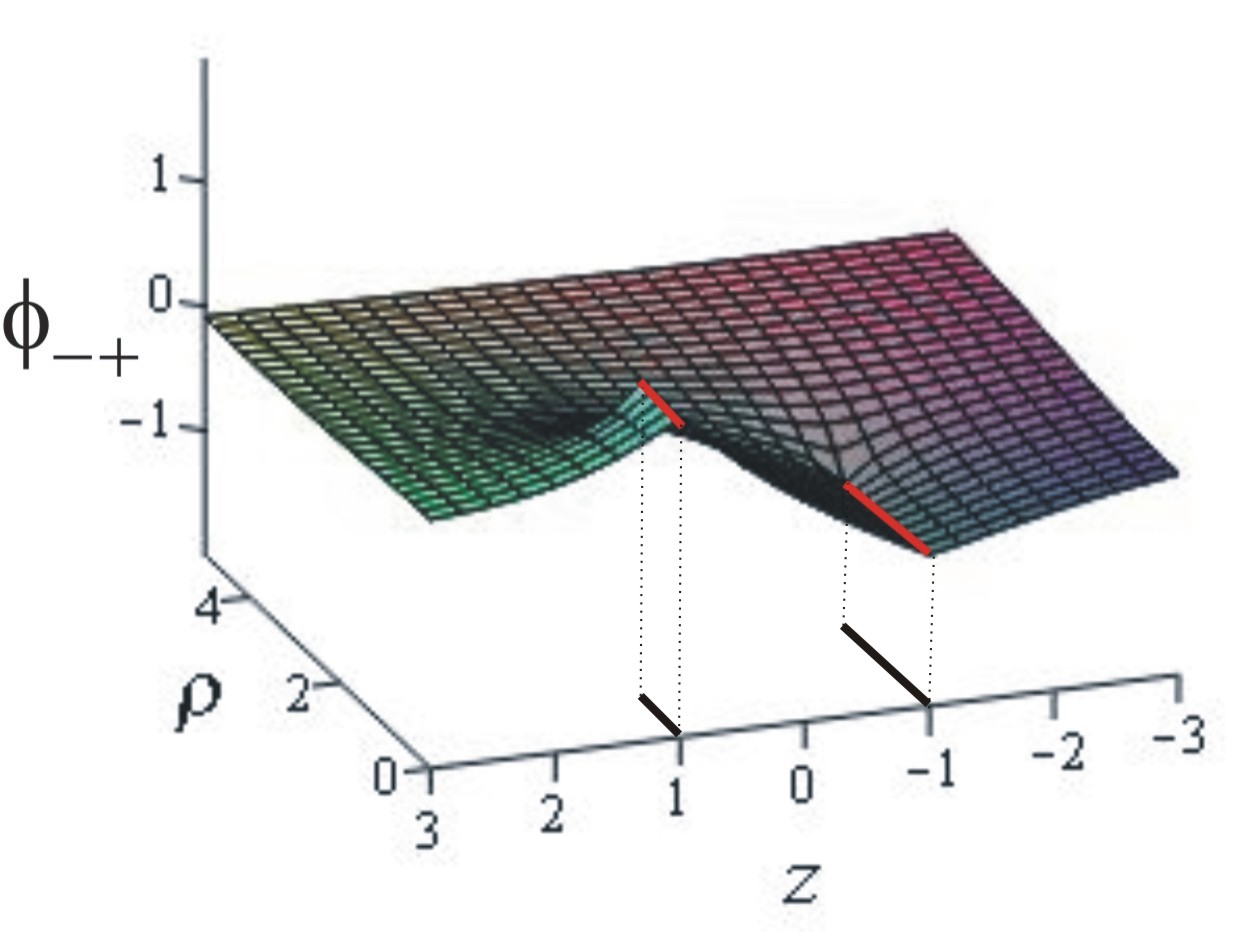}
\includegraphics[width=4.0cm]{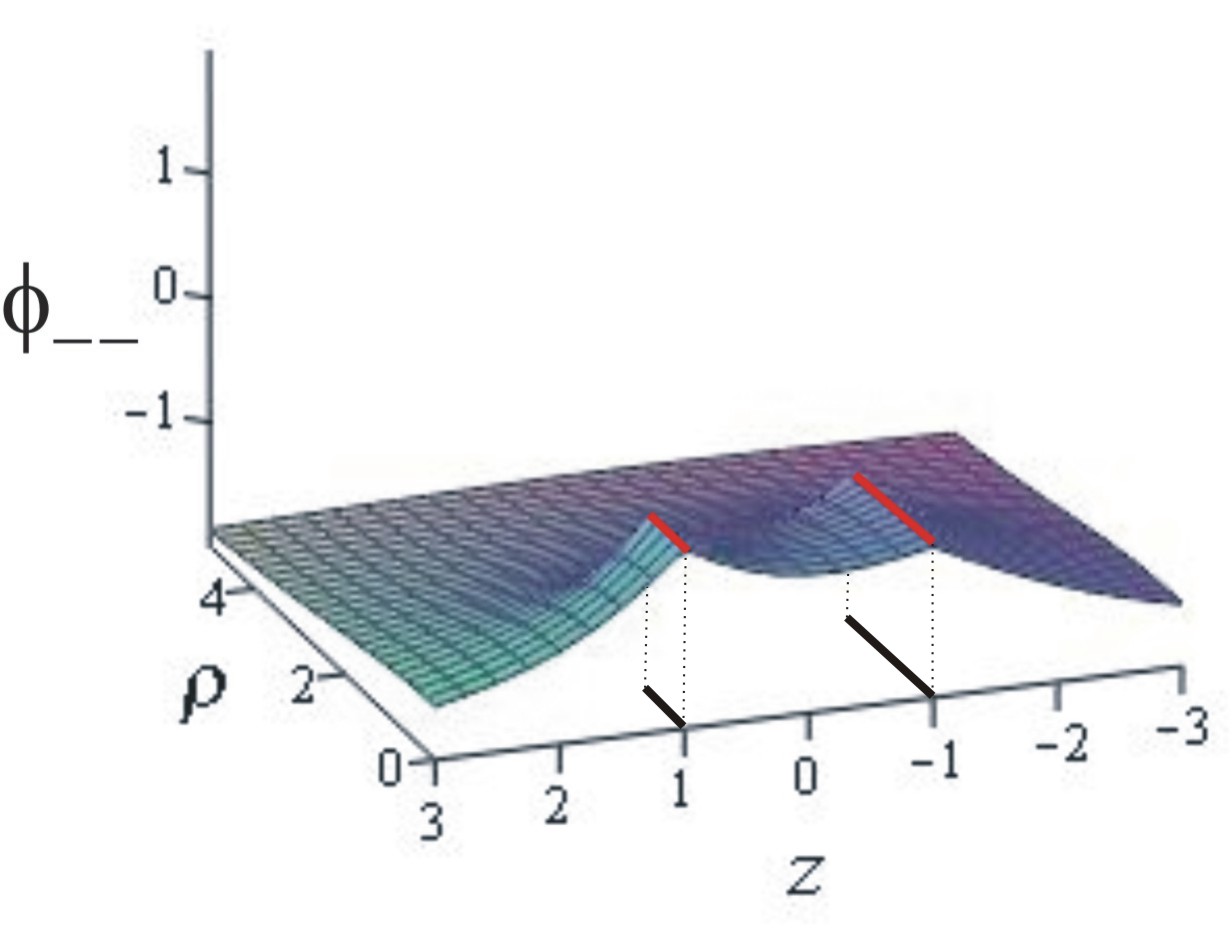}
\caption{\label{figfourphi} The plot represents separate solutions $\phi_{\epsilon_1\epsilon_2}(\rho,z)$ on half-planes ${\cal D}_{\epsilon_1\epsilon_2}$. Here $a_1=1$, $a_2=2$, and $z_1=1$, $z_2=-1$.}
\end{figure}

Note that the domain of $\phi_{\epsilon_1\epsilon_2}(\rho,z)$ is the same as that of the
single-wormhole solutions $\phi_{\epsilon_i}(\rho,z-z_i;a_i)$, i.e. the
half-plane ${\cal D}=[0,\infty)\times(-\infty,\infty)$. The segments ${\cal T}_{a_i}=\{\rho\in[0,a_i],z=z_i\}$ ($i=1,2$) on ${\cal D}$ represent two throats with radii $a_{i}$ located at $z_{i}$, respectively.
Graphs of $\phi_{\epsilon_1\epsilon_2}(\rho,z)$ are shown separately in Fig. \ref{figfourphi}.
It is seen that each of $\phi_{\epsilon_1\epsilon_2}(\rho,z)$ is not smooth
at ${\cal T}_{a_i}$.
To construct a smooth solution, we consider four identical copies of the
half-plane $\cal D$, denoted as ${\cal D}_{\epsilon_1\epsilon_2}$. On each copy ${\cal D}_{\epsilon_1\epsilon_2}$ we choose two segments ${\cal T}_{a_i}$ ($i=1,2$), and then constitute a new manifold gluing the half-planes ${\cal D}_{\epsilon_1\epsilon_2}$ along ${\cal T}_{a_i}$.
The gluing procedure is the following: ${\cal D}_{+\epsilon_2}$ and ${\cal
D}_{-\epsilon_2}$ are glued along the throat ${\cal T}_{a_1}$, and
${\cal D}_{\epsilon_1 +}$ and ${\cal D}_{\epsilon_1 -}$ are glued along the
throat ${\cal T}_{a_2}$.
The gluing procedure is schematically shown in Fig.~\ref{fig-D4}.
As the result of this topological gluing, we obtain a new manifold denoted as
$\overline{\cal D}{}_4$, consisting of four sheets ${\cal D}_{\epsilon_1\epsilon_2}$. Note that each of ${\cal D}_{\epsilon_1\epsilon_2}$ contains an asymptotically flat region $(\rho^2+z^2)^{1/2}\to\infty$, and so $\overline{\cal D}{}_4$ represents the spacetime with four asymptotically flat regions connecting by the throats ${\cal T}_{a_i}$.
It is also worth noticing that the resulting manifold $\overline{\cal D}{}_{4}$ is homeomorphic to a self-crossing four-sheeted Riemann surface.

\begin{figure}
\includegraphics[width=8cm]{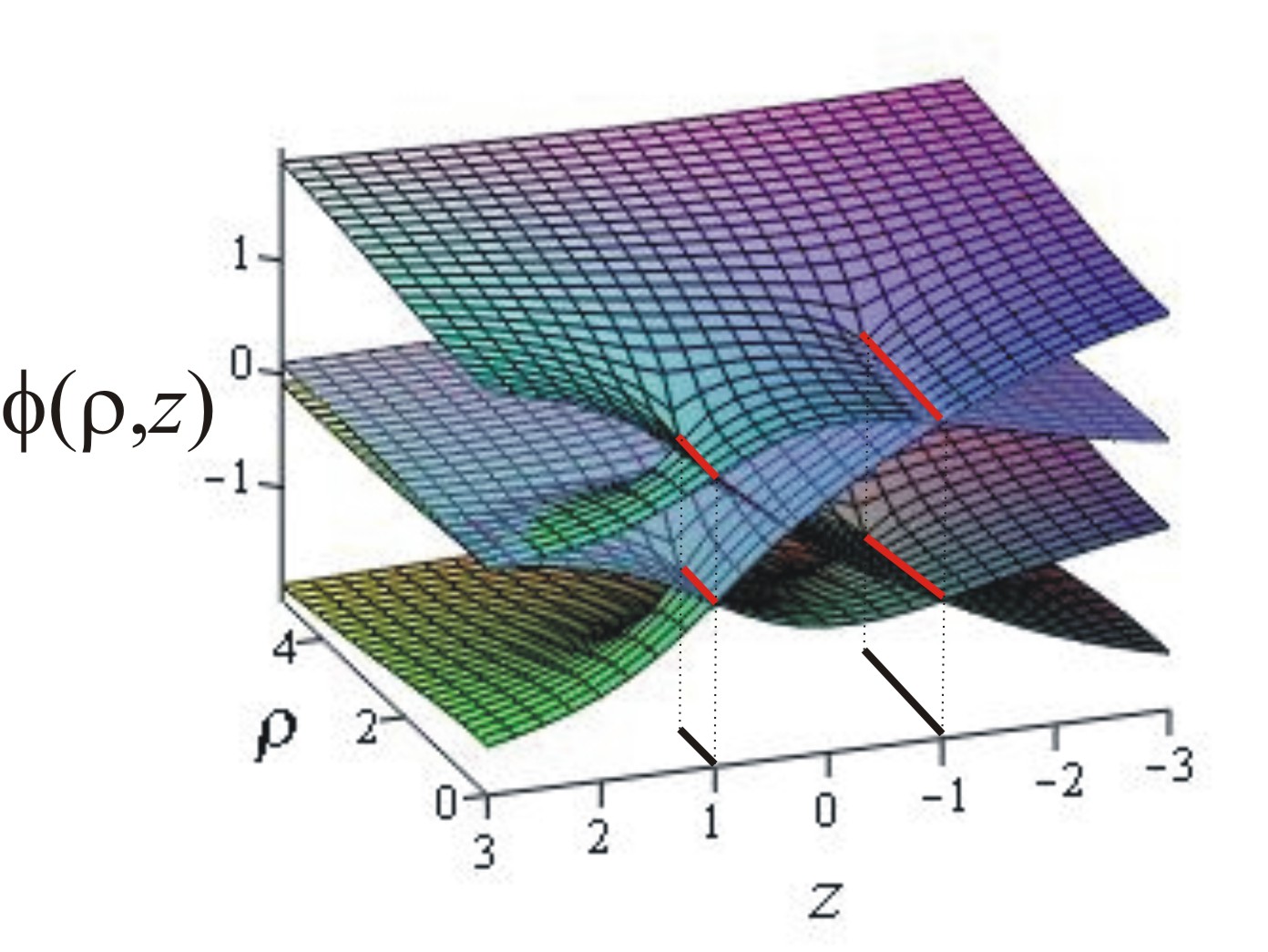}
\caption{\label{fig-phi4} The scalar field $\phi(\rho,z)$ on the manifold $\overline{\cal D}{}_{2^2}$ which possesses $2^2$ asymptotically flat regions and corresponds to the two-wormhole spacetime. Segments ${\cal T}_{a_i}=\{\rho\in[0,a_i], z_i\}$ $(i=1,2)$ correspond to the wormhole's throats with the radiuses $a_1=1$, $a_2=2$, and the positions $z_1=1$, $z_2=-1$.}
\end{figure}

The complete solution $\phi(\rho,z)$, defined on the whole manifold
$\overline{\cal D}{}_{4}$, is constituted from separate
$\phi_{\epsilon_1\epsilon_2}(\rho,z)$ as follows:
\beq\label{phi_two_throats}
\phi(\rho,z)=\bigcup_{\epsilon_1\epsilon_2}\big\{\phi_{\epsilon_1\epsilon_2}(\rho,z)~\textrm{if}~(\rho,z)\in{\cal D}_{\epsilon_1\epsilon_2}\big\}.
\eeq
To argue that $\phi(\rho,z)$ is a smooth function on $\overline{\cal
D}{}_{4}$, we note that the single-wormhole solutions
$\phi_{+}(\rho,z-z_i;a_i)$
and $\phi_{-}(\rho,z-z_i;a_i)$ are smoothly matched at the corresponding throat ${\cal T}_{a_i}=\{\rho\in[0,a_i],z=z_i\}$.
A plot of $\phi(\rho,z)$ is shown in Fig. \ref{fig-phi4}.

The function $\nu(\rho,z)$ could be found as the line integral %(\ref{nu_1wh})
$$
\nu(\rho,z)=\int_{\cal C}
\left[-{\textstyle\frac12}\rho(\phi_{,\rho}^2-\phi_{,z}^2)d\rho -\rho \phi_{,\rho}\phi_{,z}dz\right],
$$
where $\cal C$ is an arbitrary path connecting some initial point
$(\rho_0,z_0)$ and any point $(\rho,z)$ in $\overline{\cal D}{}_4$.
Note that if $(\rho_0,z_0)$ and $(\rho,z)$ are lying in different charts ${\cal
D}_{\epsilon_1\epsilon_2}$, then the path $\cal C$ should pass through some throats ${\cal T}_{a_i}$.
Coordinates of the initial point $(\rho_0,z_0)$ are determined by boundary conditions for the function $\nu$. Imposing the following boundary conditions
$$
\lim\limits_{\rho\to\infty,z\to -\infty}\nu(\rho,z)=0,\quad (\rho,z)\in {\cal D_{++}},
$$
substituting $\phi$ from Eq. \Ref{phi_two_throats}, and integrating, we can obtain the set of functions $\nu_{\epsilon_1\epsilon_2}(\rho,z)$ defined on ${\cal
D}_{\epsilon_1\epsilon_2}$, respectively.
Finally, the complete solution $\nu(\rho,z)$ defined on $\overline{\cal D}{}_4$ is constituted from the separate $\nu_{\epsilon_1\epsilon_2}(\rho,z)$ as follows
\beq\label{sol-nu1}
\nu(\rho,z)=\bigcup_{\epsilon_1, \epsilon_2}\left\{
\nu_{\epsilon_1\epsilon_2}(\rho,z) ~\textrm{if }~ (\rho,z)\in {\cal D}_{\epsilon_1\epsilon_2}
\right\}.
\eeq

%%%%%%%%%%%%%%%%%%%%%%%%%%%%%%%%%%%%%%%%%%%%%%%%%%%%
\subsection{$N$ wormholes}
%%%%%%%%%%%%%%%%%%%%%%%%%%%%%%%%%%%%%%%%%%%%%%%%%%%%
A multi-wormhole solution could be constructed analogously to the two-wormhole one.
For this aim we consider $2^N$ different superpositions of $N$
single-wormhole solutions:
\beq\label{phiN}
\phi_{\epsilon_1\epsilon_2\dots\epsilon_N}(\rho,z)=\frac1N\sum_{i=1}^{N}
\phi_{\epsilon_i }(\rho, z-z_i;a_i),
\eeq
where $i=1,2,\dots,N$, and $\epsilon_i$ is an index which stands on the $i$th position
and takes two values, $+$ or $-$, i.e. $\epsilon_i=\pm$.
Each of $\phi_{\epsilon_1\epsilon_2\dots\epsilon_N}(\rho,z)$ is defined on the
half-plane ${\cal D}=[0,\infty)\times(-\infty,\infty)$. Consider $2^N$
identical copies of $\cal D$, denoted as ${\cal
D}_{\epsilon_1\epsilon_2\dots\epsilon_N}$, and then construct a new
manifold $\overline{\cal D}{}_{2^N}$ gluing the half-planes ${\cal
D}_{\epsilon_1\epsilon_2\dots\epsilon_N}$. The gluing procedure is the
following: the charts ${\cal
D}_{\epsilon_1\cdots\{\epsilon_k=+\}\cdots\epsilon_N}$ and ${\cal
D}_{\epsilon_1\cdots\{\epsilon_k=-\}\cdots\epsilon_N}$ are glued along the
$k$th segment $\{\rho\in[0,a_k],z=z_k\}$.

As the result of this topological gluing, we obtain a new manifold $\overline{\cal D}{}_{2^N}$,
consisting of $2^N$ sheets ${\cal
D}_{\epsilon_1\epsilon_2\dots\epsilon_N}$. Each of ${\cal
D}_{\epsilon_1\epsilon_2\dots\epsilon_N}$ contains an asymptotically flat region $(\rho^2+z^2)^{1/2}\to\infty$, and so $\overline{\cal D}{}_{2^N}$ represents the spacetime with $2^N$ asymptotically flat regions connecting by the throats ${\cal T}_{a_i}$.
It is also worth noticing that the resulting manifold $\overline{\cal D}{}_{2^N}$ is homeomorphic to a self-crossing $2^N$-sheeted Riemann surface.

The complete solution $\phi(\rho,z)$, defined on the whole manifold $\overline{\cal D}{}_{2^N}$,
is constituted from separate $\phi_{\epsilon_1\epsilon_2\dots\epsilon_N}(\rho,z)$ as follows:
\beq\label{phi_N_throats}
\phi(\rho,z)=\bigcup_{\epsilon_1\epsilon_2\dots\epsilon_N}\big\{\phi_{\epsilon_1\epsilon_2\dots\epsilon_N}(\rho,z)~\textrm{if
}~(\rho,z)\in{\cal D}_{\epsilon_1\epsilon_2\dots\epsilon_N}\big\}.
\eeq

The metric function $\nu(\rho,z)$ is found as the line integral
$$
\nu(\rho,z)=\int_{\cal C}
\left[-{\textstyle\frac12}\rho(\phi_{,\rho}^2-\phi_{,z}^2)d\rho -\rho \phi_{,\rho}\phi_{,z}dz\right],
$$
where $\phi$ is given by Eq. \Ref{phi_N_throats}, and $\cal C$ is an arbitrary path connecting some initial point $(\rho_0,z_0)$ and any point $(\rho,z)$ in $\overline{\cal D}{}_{2^N}$.

\begin{figure}
\includegraphics[width=8cm]{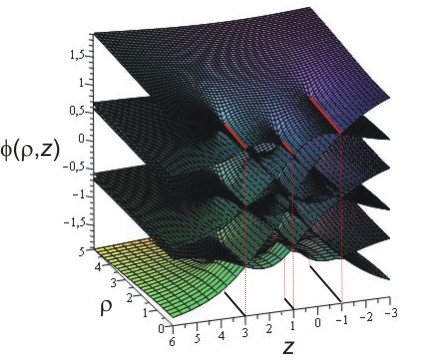}
\caption{\label{fig-phi3} The scalar field $\phi(\rho,z)$ on the manifold $\overline{\cal D}{}_{2^3}$ which possesses $2^3$ asymptotically flat regions and corresponds to the three-wormhole spacetime. Segments ${\cal T}_{a_i}=\{\rho\in[0,a_i], z_i\}$ $(i=1,2,3)$ correspond to the wormhole's throats with the radiuses $a_1=1$, $a_2=1.5$, $a_3=2$, and the positions $z_1=1$, $z_2=3$, $z_3=-1$.}
\end{figure}

As an illustration of a multi-wormhole configuration with $N>2$, in Fig. \ref{fig-phi3} we depict
the scalar field in the wormhole spacetime $\overline{\cal D}{}_{2^3}$ with $2^3$ asymptotically flat regions connecting by the three throats.

\section{Conclusions}
Many years ago, in 1921 Bach and Weyl \cite{BachWeyl} derived the method of superposition to construct new axially symmetric vacuum solutions of General Relativity and obtained the famous solution interpreted as equilibrium configurations of a pair of Schwarzschild black holes. In this paper we have extended the Bach-Weyl approach to non-vacuum configurations with massless scalar fields. Considering phantom scalar fields with the negative kinetic energy, we have constructed  multi-wormhole solutions describing an axially symmetric superposition of $N$ wormholes. Let us enumerate basic properties of the obtained solutions.
\begin{itemize}
\item
The most unusual property is that the multi-wormhole spacetime has a complicated topological structure. Namely, in the spacetime there exist $2^N$ asymptotically flat regions connected by throats, so that the resulting spacetime manifold is homeomorphic to a self-crossing $2^N$-sheeted Riemann surface.

\item
The spacetime of multi-wormholes is everywhere regular and has no event horizons.
This feature drastically tells the multi-wormhole configuration from other axially symmetric vacuum solutions. As is known, all static solutions, representing the superposition of two or more axially symmetric bodies, inevitably contain gravitationally inert singular structures, `struts' and `membranes', that keep the two bodies apart making a stable configuration. This is an expected ``price to pay'' for the stationarity. However, the multi-wormholes are static without any singular struts. Instead, the stationarity of the multi-wormhole configuration is provided by the phantom scalar field with the negative kinetic energy. Phantom scalars represent exotic matter violating the null energy condition. Therefore, now an ``exoticism'' of the matter source supporting the multi-wormholes is a price for the stationarity.

%\item
%The multi-wormholes have been constructed with using the phantom scalar field. Therefore, as expected, the null energy condition is violated in the multi-wormhole spacetime.
\end{itemize}

%The stability of wormhole configurations is an important test of their possible viability.
%The stability of wormholes supported by phantom scalar fields was intensively investigated
%in the literature [21-24], and the final resolution states that both static and non-static (see
%Ref.   [21])  scalar  wormholes  are  unstable

%----------------------------------------------------------------
\section*{Acknowledgments}
%----------------------------------------------------------------
The work was supported by the Russian Government Program of Competitive Growth of Kazan Federal University and, partially, by the Russian Foundation for Basic Research grants No. 14-02-00598 and 15-52-05045.

%----------------------------------------------------------------

\end{document}